\documentclass[aps,prd,epsfig,preprint,amsmath,amssymb,superscriptaddress,showpacs,nofootinbib]{revtex4}
\usepackage{multirow}
\usepackage{graphicx,bm}
\usepackage[usenames]{color}
\usepackage{float}
\usepackage{caption}
\usepackage{subcaption}
\usepackage{slashed}
\begin{document}
\draft
\title{Search for the anomalous $ZZ\gamma$ and $Z\gamma\gamma$ couplings via $\nu \nu \gamma$ production at the CLIC}

\author{S. Spor}
\email[]{serdar.spor@beun.edu.tr}
\affiliation{Department of Medical Imaging Techniques, Zonguldak B\"{u}lent Ecevit University, 67100, Zonguldak, T\"{u}rkiye.}

\author{E. Gurkanli}
\email[]{egurkanli@sinop.edu.tr}
\affiliation{ Department of Physics, Sinop University, 57000, T\"{u}rkiye.}

\author{M. K{\"o}ksal}
\email[]{mkoksal@cumhuriyet.edu.tr} 
\affiliation{Department of Physics, Sivas Cumhuriyet University, 58140, Sivas, T\"{u}rkiye.} 

\date{\today}

\begin{abstract}
The non-Abelian gauge structure of the Standard Model implies the presence of the multi-boson self-interactions. Precise measurements in experimental and theoretical studies of these interactions allow not only testing the nature of the Standard Model but also new physics contribution coming from the beyond Standard Model. These interactions can be examined using a model-independent way in effective theory approach that composes the motivation part of this study. In this paper, we examine the anomalous $ZZ\gamma$ and $Z\gamma\gamma$ neutral triple gauge couplings via the process $e^{-} e^{+}\to Z\gamma$ for the neutrino-antineutrino pair decay of $Z$ boson. It has performed with both unpolarized and polarized electron beams at the Compact Linear Collider with $\sqrt{s}= 3$ TeV. The study focused on $CP$-conserving $C_{\widetilde{B}W}/{\Lambda^4}$ and $CP$-violating $C_{BB}/{\Lambda^4}$, $C_{BW}/{\Lambda^4}$, $C_{WW}/{\Lambda^4}$ couplings. Obtained sensitivities on the anomalous neutral triple gauge couplings with $95\%$ Confidence Level are given with systematic uncertainties of $0\%$, $5\%$ and $10\%$ for unpolarized, $-80\%$ and $80\%$ polarized electron beams with integrated luminosities of ${\cal L}_{\text{int}}=5$ $\rm ab^{-1}$, ${\cal L}_{\text{int}}=4$ $\rm ab^{-1}$ and ${\cal L}_{\text{int}}=1$ $\rm ab^{-1}$, respectively. Comparing the latest experimental limits and related phenomenological studies, our results on the anomalous neutral gauge couplings are set more stringent sensitivity between 10-30 times of magnitude. 

\end{abstract}


\maketitle

\section{Introduction}

The gauge boson self-interactions are described by the non-Abelian $SU(2)_L \times U(1)_Y$ gauge theory of the Standard Model (SM). These self-interactions can be defined by triple gauge boson $WWV$, $ZV\gamma$ and $ZZV$ $(V=\gamma$, $Z)$ couplings \cite{Baur:2000hfg}. However, the couplings with the $Z$ boson and the photon are not found at the tree-level in the SM, since the $Z$ boson has no electrical charge. The goal of detecting any deviations from the SM predictions that are popular nowadays is to explore new physics beyond the SM. Anomalous neutral triple gauge couplings (aNTGC) between the photon and the $Z$ boson $(Z\gamma\gamma$ and $ZZ\gamma)$ are of unique importance in the investigation of new physics beyond the SM. Because of the absence of triple gauge couplings between the photon and the $Z$ boson in the SM, the deviation from the SM prediction in the presence of $Z\gamma\gamma$ and $ZZ\gamma$ vertices is sensitive evidence for the new physics. 

It has been widely studied the $ZZ$ and $Z\gamma$ production in $e^-e^+$ colliders \cite{Choudhury:1994ywq,Atag:2004ybz,Ots:2004twm,Ots:2006gsd,Rodriguez:2009rnw,Ananthanarayan:2012onz,Ananthanarayan:2014cal,Rahaman:2016nzs,Rahaman:2017qed,Ellis:2020ekm,Fu:2021jec,Ellis:2021rop} and $pp$ colliders \cite{Baur:1993fkx,Senol:2018gvg,Rahaman:2019tnp,Senol:2019ybv,Senol:2020hbh,Yilmaz:2020ser,Yilmaz:2021dbm,Hernandez:2021wsz,Biekotter:2021ysx,Yang:2021tgw} to investigate aNTGC. In addition, it is easier to experimentally detect final states in vector boson pair production processes, and clean signatures at the detector and rich data based on a higher signal-to-background ratio are obtained from the interactions of these processes \cite{Geng:2019ebo,Gounaris:2003lsd}.

The Effective Field Theory (EFT) approach has been used to parameterize these new physics effects with higher-dimension operators in a model-independent way. The method used to study the aNTGC with the SM gauge group is the following: add high-dimensional operators that are invariant under $SU(2)_L \times U(1)_Y$ to the SM Lagrangian and acquire the effective vertices with the anomalous couplings after electroweak symmetry breaking \cite{Rahaman:2016nzs}. The effective Lagrangian of the EFT which includes SM interactions and dimension-eight operators describing aNTGC can be written as \cite{Degrande:2014ydn}

\begin{eqnarray}
\label{eq.1} 
{\cal L}^{\text{nTGC}}={\cal L}_{\text{SM}}+\sum_{i}\frac{C_i}{\Lambda^{4}}({\cal O}_i+{\cal O}_i^\dagger)
\end{eqnarray}

{\raggedright where $\Lambda$ is the new physics scale and the index $i$ labels the four operators given below}

\begin{eqnarray}
\label{eq.2} 
{\cal O}_{\widetilde{B}W}=iH^{\dagger} \widetilde{B}_{\mu\nu}W^{\mu\rho} \{D_\rho,D^\nu \}H,
\end{eqnarray}
\begin{eqnarray}
\label{eq.3} 
{\cal O}_{BW}=iH^\dagger B_{\mu\nu}W^{\mu\rho} \{D_\rho,D^\nu \}H,
\end{eqnarray}
\begin{eqnarray}
\label{eq.4} 
{\cal O}_{WW}=iH^\dagger W_{\mu\nu}W^{\mu\rho} \{D_\rho,D^\nu \}H,
\end{eqnarray}
\begin{eqnarray}
\label{eq.5} 
{\cal O}_{BB}=iH^\dagger B_{\mu\nu}B^{\mu\rho} \{D_\rho,D^\nu \}H.
\end{eqnarray}

Here, $B_{\mu\nu}$ and $W^{\mu\nu}$ are the field strength tensors and $D_\mu$ is the covariant derivative. The first operator is $CP$-even and the last three operators are $CP$-odd. They are used in the definitions of operators as follows:

\begin{eqnarray}
\label{eq.6} 
B_{\mu\nu}=\left(\partial_\mu B_\nu - \partial_\nu B_\mu\right),
\end{eqnarray}
\begin{eqnarray}
\label{eq.7} 
W_{\mu\nu}=\sigma^i\left(\partial_\mu W_\nu^i - \partial_\nu W_\mu^i + g\epsilon_{ijk}W_\mu^j W_\nu^k\right),
\end{eqnarray}

{\raggedright with $\langle \sigma^i\sigma^j\rangle=\delta^{ij}/2$ and}

\begin{eqnarray}
\label{eq.8} 
D_\mu \equiv \partial_\mu - i\frac{g^\prime}{2}B_\mu Y - ig_W W_\mu^i\sigma^i.
\end{eqnarray}

When the new physics energy scale is high, the biggest new physics contribution to $Z\gamma$ pair generation comes from the interference between the SM and the dimension-eight operators. As long as the interferences between the SM and the dimension-eight and dimension-ten operators are not both suppressed, only dimension-eight operator ${\cal O}({\Lambda^{-8}})$ does not add up a contribution above the new physics with high-energy scale. The dimension-six operators do not have any effect on aNTGC at tree-level but the order ${\alpha \hat{s}}/{4\pi\Lambda^2}$ effects can occur on aNTGC at one-loop. The contributions of the dimension-eight operators at tree-level are of the order ${\upsilon^2\hat{s}}/{\Lambda^4}$. The contribution of the dimension-eight operators is more forefront than one-loop contribution of the dimension-six operator with $\Lambda \lesssim \sqrt{4\pi\hat{s}/\alpha}$ \cite{Degrande:2014ydn}.

Effective Lagrangian for aNTGC with dimension-six and dimension-eight operators is given by \cite{Gounaris:2000svs}

\begin{eqnarray}
\label{eq.9} 
\begin{split}
{\cal L}_{\text{aNTGC}}^{\text{dim-6,8}}=&\frac{g_e}{m_Z^2}\Bigg[-[f_4^\gamma(\partial_\mu F^{\mu\beta})+f_4^Z(\partial_\mu Z^{\mu\beta})]Z_\alpha (\partial^\alpha Z_\beta)+[f_5^\gamma(\partial^\sigma F_{\sigma\mu})+f_5^Z (\partial^\sigma Z_{\sigma\mu})]\widetilde{Z}^{\mu\beta}Z_\beta  \\
&-[h_1^\gamma (\partial^\sigma F_{\sigma\mu})+h_1^Z (\partial^\sigma Z_{\sigma\mu})]Z_\beta F^{\mu\beta}-[h_3^\gamma(\partial_\sigma F^{\sigma\rho})+h_3^Z(\partial_\sigma Z^{\sigma\rho})]Z^\alpha \widetilde{F}_{\rho\alpha}   \\
&-\bigg\{\frac{h_2^\gamma}{m_Z^2}[\partial_\alpha \partial_\beta \partial^\rho F_{\rho\mu}]+\frac{h_2^Z}{m_Z^2}[\partial_\alpha \partial_\beta(\square+m_Z^2)Z_\mu]\bigg\}Z^\alpha F^{\mu\beta}   \\
&+\bigg\{\frac{h_4^\gamma}{2m_Z^2}[\square\partial^\sigma F^{\rho\alpha}]+\frac{h_4^Z}{2m_Z^2}[(\square+m_Z^2)\partial^\sigma Z^{\rho\alpha}]\bigg\}Z_\sigma\widetilde{F}_{\rho\alpha}\Bigg],
\end{split}
\end{eqnarray}

{\raggedright where $\widetilde{Z}_{\mu\nu}=1/2\epsilon_{\mu\nu\rho\sigma}Z^{\rho\sigma}$ $(\epsilon^{0123}=+1)$ with field strength tensor $Z_{\mu\nu}=\partial_\mu Z_\nu - \partial_\nu Z_\mu$ and similarly for the electromagnetic field tensor $F_{\mu\nu}$. However, $f_4^V$, $h_1^V$, $h_2^V$ are three $CP$-violating couplings while $f_5^V$, $h_3^V$, $h_4^V$ are three $CP$-conserving couplings $(V=\gamma$, $Z)$. In the SM at tree-level, all couplings are zero. In the Lagrangian, the couplings $h_2^V$ and $h_4^V$ correspond to dimension-eight and the other four couplings to dimension-six.}

The couplings in the effective Lagrangian in Eq.~(\ref{eq.9}) are related to the couplings of the operators in Eqs.~(\ref{eq.2}-\ref{eq.5}) when $SU(2)_L \times U(1)_Y$ gauge invariance is considered \cite{Rahaman:2020fdf}. The $CP$-conserving anomalous couplings with two on-shell $Z$ bosons and one off-shell $V=\gamma$ or $Z$ boson in the anomalous vertex for the neutral gauge bosons are written by \cite{Degrande:2014ydn}

\begin{eqnarray}
\label{eq.10} 
f_5^Z=0,
\end{eqnarray}
\begin{eqnarray}
\label{eq.11} 
f_5^\gamma=\frac{\upsilon^2 m_Z^2}{4c_\omega s_\omega} \frac{C_{\widetilde{B}W}}{\Lambda^4}
\end{eqnarray}

{\raggedright and the $CP$-violating anomalous couplings by}

\begin{eqnarray}
\label{eq.12} 
f_4^Z=\frac{m_Z^2 \upsilon^2 \left(c_\omega^2 \frac{C_{WW}}{\Lambda^4}+2c_\omega s_\omega \frac{C_{BW}}{\Lambda^4}+4s_\omega^2 \frac{C_{BB}}{\Lambda^4}\right)}{2c_\omega s_\omega},
\end{eqnarray}
\begin{eqnarray}
\label{eq.13} 
f_4^\gamma=-\frac{m_Z^2 \upsilon^2 \left(-c_\omega s_\omega \frac{C_{WW}}{\Lambda^4}+\frac{C_{BW}}{\Lambda^4}(c_\omega^2-s_\omega^2)+4c_\omega s_\omega \frac{C_{BB}}{\Lambda^4}\right)}{4c_\omega s_\omega}.
\end{eqnarray}

The $CP$-conserving anomalous couplings with one on-shell $Z$ boson, one on-shell photon and one off-shell $V=\gamma$ or $Z$ boson in the anomalous vertex for the neutral gauge bosons are written by \cite{Degrande:2014ydn}

\begin{eqnarray}
\label{eq.14} 
h_3^Z=\frac{\upsilon^2 m_Z^2}{4c_\omega s_\omega} \frac{C_{\widetilde{B}W}}{\Lambda^4},
\end{eqnarray}
\begin{eqnarray}
\label{eq.15} 
h_4^Z=h_3^\gamma=h_4^\gamma=0 ,
\end{eqnarray}

{\raggedright and the $CP$-violating anomalous couplings by}

\begin{eqnarray}
\label{eq.16} 
h_1^Z=\frac{m_Z^2 \upsilon^2 \left(-c_\omega s_\omega \frac{C_{WW}}{\Lambda^4}+\frac{C_{BW}}{\Lambda^4}(c_\omega^2-s_\omega^2)+4c_\omega s_\omega \frac{C_{BB}}{\Lambda^4}\right)}{4c_\omega s_\omega},
\end{eqnarray}
\begin{eqnarray}
\label{eq.17} 
h_2^Z=h_2^\gamma=0,
\end{eqnarray}
\begin{eqnarray}
\label{eq.18} 
h_1^\gamma=-\frac{m_Z^2 \upsilon^2 \left(s_\omega^2 \frac{C_{WW}}{\Lambda^4}-2c_\omega s_\omega \frac{C_{BW}}{\Lambda^4}+4c_\omega^2 \frac{C_{BB}}{\Lambda^4}\right)}{4c_\omega s_\omega}.
\end{eqnarray}

Many couplings given above are equal to zero, due to the disappearance of the $CP$-conserving $Z\gamma\gamma$ and $ZZZ$ vertices. These four dimension-eight coefficients describe aNTGC in Eqs.~(\ref{eq.11}-\ref{eq.14},\ref{eq.16},\ref{eq.18}) are $CP$-conserving $C_{\widetilde{B}W}/{\Lambda^4}$ and $CP$-violating $C_{BB}/{\Lambda^4}$, $C_{BW}/{\Lambda^4}$, $C_{WW}/{\Lambda^4}$. We have investigated the sensitivity on aNTGC, $ZZ\gamma$ and $Z\gamma\gamma$, with dimension-eight couplings $C_{\widetilde{B}W}/{\Lambda^4}$, $C_{BB}/{\Lambda^4}$, $C_{BW}/{\Lambda^4}$ and $C_{WW}/{\Lambda^4}$ via the process $e^-e^+\,\rightarrow\,Z\gamma$ at the Compact Linear Collider (CLIC), which is designed with the center-of-mass energy of 3 TeV. The latest experimental limits on dimension-eight couplings are studied through process $pp\rightarrow Z\gamma \rightarrow \nu\bar{\nu}\gamma$ at center-of-mass energy of 13 TeV with integrated luminosity of 36.1 fb$^{-1}$ at the CERN LHC \cite{Aaboud:2018ybz}. These 95$\%$ Confidence Level (C.L.) experimental limits are given as

\begin{eqnarray}
\label{eq.19} 
-1.1\, \text{TeV}^{-4}<\frac{C_{\widetilde{B}W}}{\Lambda^4}<1.1 \, \text{TeV}^{-4},
\end{eqnarray}
\begin{eqnarray}
\label{eq.20} 
-2.3\, \text{TeV}^{-4}<\frac{C_{WW}}{\Lambda^4}<2.3 \, \text{TeV}^{-4},
\end{eqnarray}
\begin{eqnarray}
\label{eq.21} 
-0.65\, \text{TeV}^{-4}<\frac{C_{BW}}{\Lambda^4}<0.64 \, \text{TeV}^{-4},
\end{eqnarray}
\begin{eqnarray}
\label{eq.22} 
-0.24\, \text{TeV}^{-4}<\frac{C_{BB}}{\Lambda^4}<0.24 \, \text{TeV}^{-4}.
\end{eqnarray}

\section{Cross-sections and Events} \label{Sec2}

\begin{figure}[t]
\centerline{\scalebox{0.09}{\includegraphics{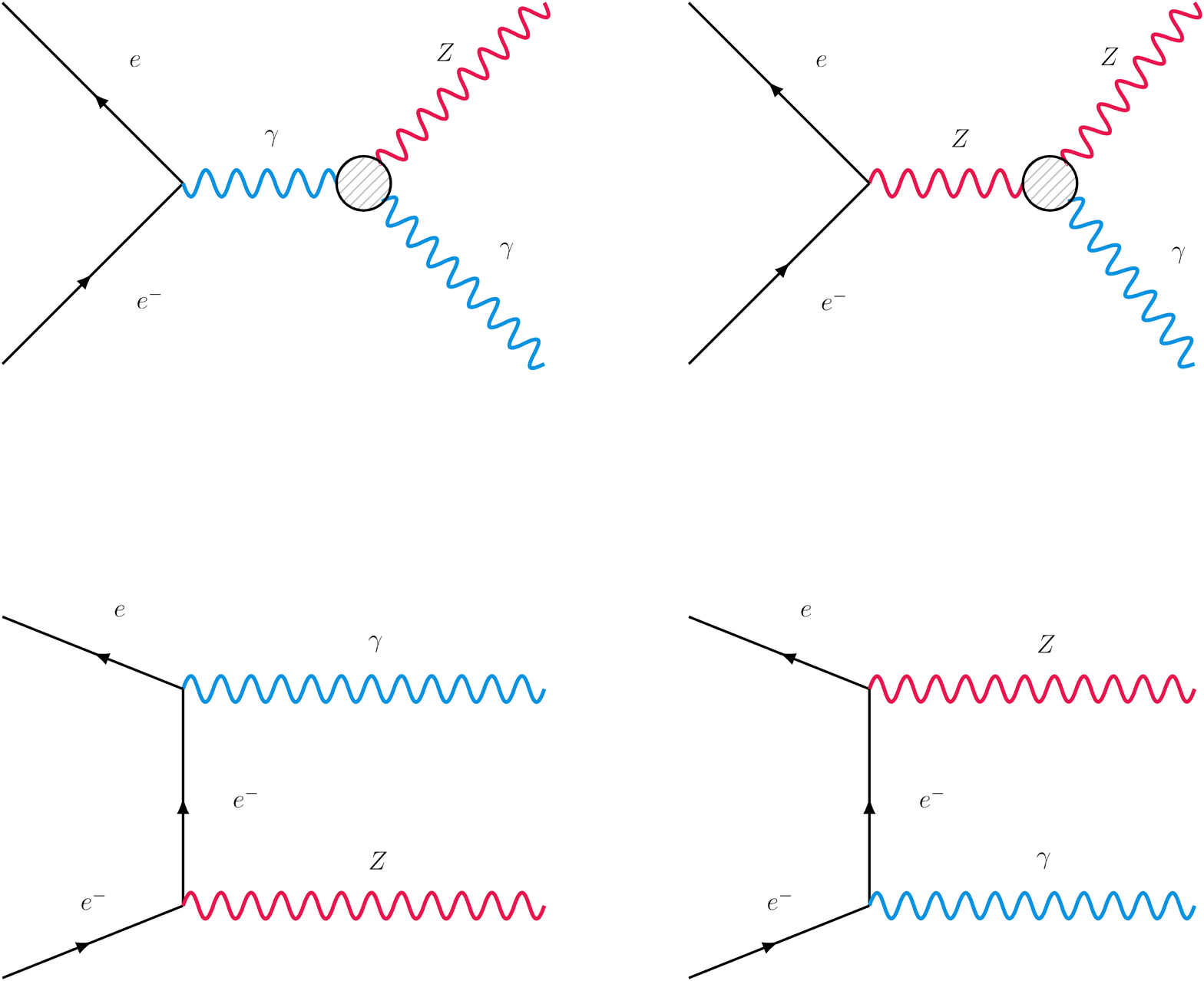}}}
\caption{Feynman diagrams of the process $e^- e^{+} \to Z\gamma $ including the anomalous contribution of $ZZ\gamma$ and $Z\gamma\gamma$ vertices and the SM contribution.}
\label{Fig.1}
\end{figure}

The Feynman diagrams for the process $e^-e^+\,\rightarrow\,Z\gamma$ are given in Fig.~\ref{Fig.1}. Here, the first two Feynman diagrams contain contributions of new physics beyond the SM from the anomalous $ZZ\gamma$ and $Z\gamma\gamma$ couplings, while the last two Feynman diagrams contain SM contributions. In this study, the $\nu\bar{\nu}\gamma$ final state in the $Z\gamma$ production process is discussed. The process involving the decay of the $Z$ boson into neutrinos, $Z(\nu\bar{\nu})\gamma$, has many advantages over processes that involve decay into hadrons $Z(q\bar{q})\gamma$ or charged leptons $Z(\ell^-\ell^+)\gamma$. In the final state, the hadron channel does not have clean data due to the large number of multijet backgrounds. Besides, the fact that neutrino pair decay has a higher $Z$ boson branching ratio than charged leptons provides the opportunity to study $Z\gamma$ production in the more energetic region where the sensitivity is high \cite{Aaboud:2018ybz}.

SM background processes that have the same or similar final state topology with $\nu\bar{\nu}\gamma$ final state investigated in the $e^-e^+\,\rightarrow\,\nu\bar{\nu}\gamma$ signal process are considered. The main background, denoted by SM, is the SM background, which has the same final state as the signal process and contains contributions from the last two Feynman diagrams in Fig.~\ref{Fig.1}. In addition to the SM background, we consider both the $W$ boson pair $(e^-e^+\,\rightarrow\,W^-W^+\gamma)$ and the top-antitop quark pair $(e^-e^+\,\rightarrow\,t\bar{t}\gamma)$, along with a photon as relevant backgrounds. The $W^-W^+\gamma$ is considered as background due to having two pairs of a charged lepton and a neutrino when leptonic decay of $W$ bosons. The $t\bar{t}\gamma$ process is also added as an another background since there are two b-jets and two $W$ bosons as a result of the decay of each top quark to $W^\pm b$. When detailed, if two b-jets are any misidentification of the light quark by being mistagged in detector and two $W$ bosons decay leptonically, this background process has also a similar final state topology as the signal process.

\begin{table}
\caption{Particle-level selections cuts for the $Z\gamma$ signal at the CLIC.}
\label{tab1}
\begin{tabular}{|c|c|c}
\hline
Kinematic cuts & $C_{BB}/\Lambda^{4}$, $C_{BW}/\Lambda^{4}$, $C_{\widetilde{B}W}/\Lambda^{4}$, $C_{WW}/\Lambda^{4}$ \\
\hline
\hline
Cut-1   &  $|\eta^{\gamma}| < 2.5$   \\
\hline
Cut-2   & $\slashed{E}_T > 300$ GeV \\
\hline
Cut-3   & $p^\gamma_T > 300$ GeV\\
\hline
\end{tabular}
\end{table}

\begin{figure}[t]
\centerline{\scalebox{0.9}{\includegraphics{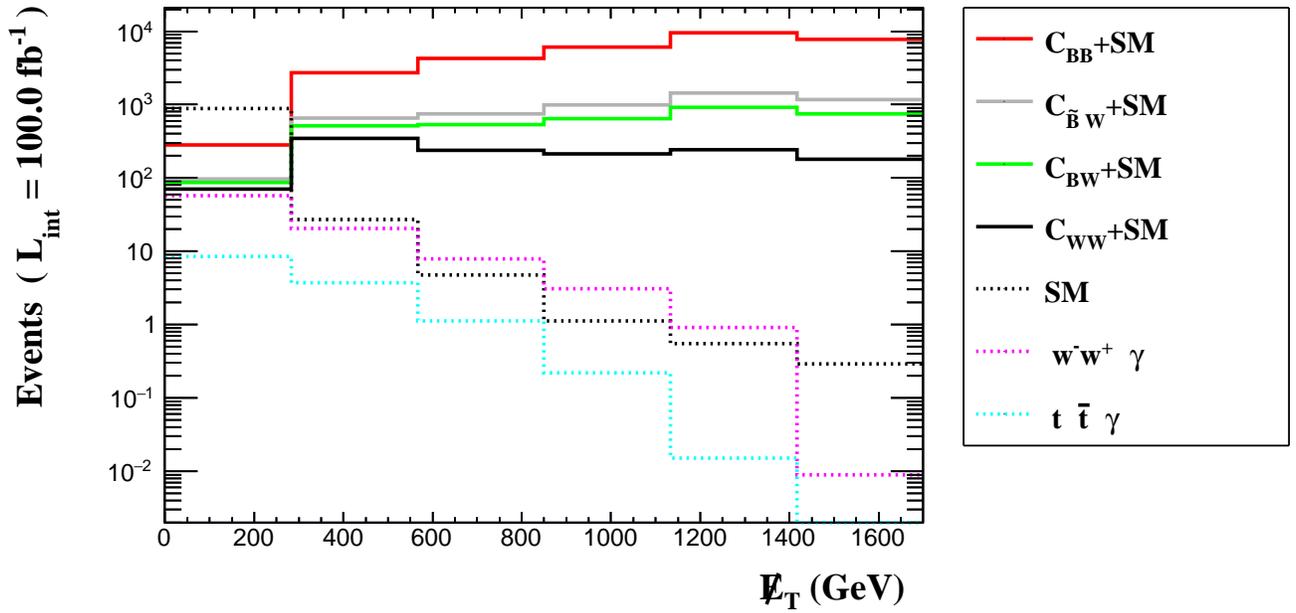}}}
\caption{The number of expected events as a function of $\slashed{E}_T$ missing energy transverse for the $e^+e^- \to Z\gamma$ signal and relevant backgrounds at the CLIC. In this figure, we have taken a value of  $1\hspace{1mm}{\rm TeV}^{-4}$ for each anomalous coupling.}
\label{Fig.2}
\end{figure}

\begin{figure}[t]
\centerline{\scalebox{0.9}{\includegraphics{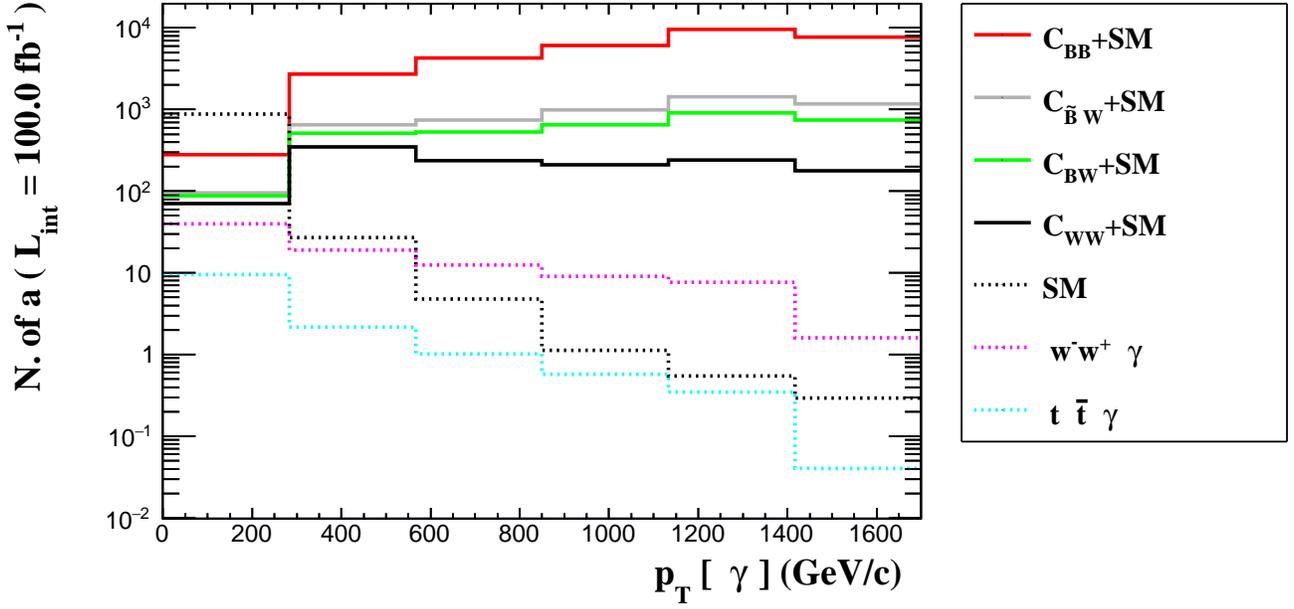}}}
\caption{The number of expected events as a function of $p^{\gamma}_T$ for the $e^+e^- \to Z\gamma$ signal and backgrounds at the CLIC with $\sqrt{s}=3$ TeV and $P_{e^-}=0\%$. The distributions are for $C_{BB}/\Lambda^{4}$, $C_{BW}/\Lambda^{4}$, $C_{\widetilde{B}W}/\Lambda^{4}$, $C_{WW}/\Lambda^{4}$ and relevant backgrounds. In this figure, we have taken a value of  $1\hspace{1mm}{\rm TeV}^{-4}$ for each anomalous coupling.}
\label{Fig.3}
\end{figure}

\begin{figure}[t]
\centerline{\scalebox{0.9}{\includegraphics{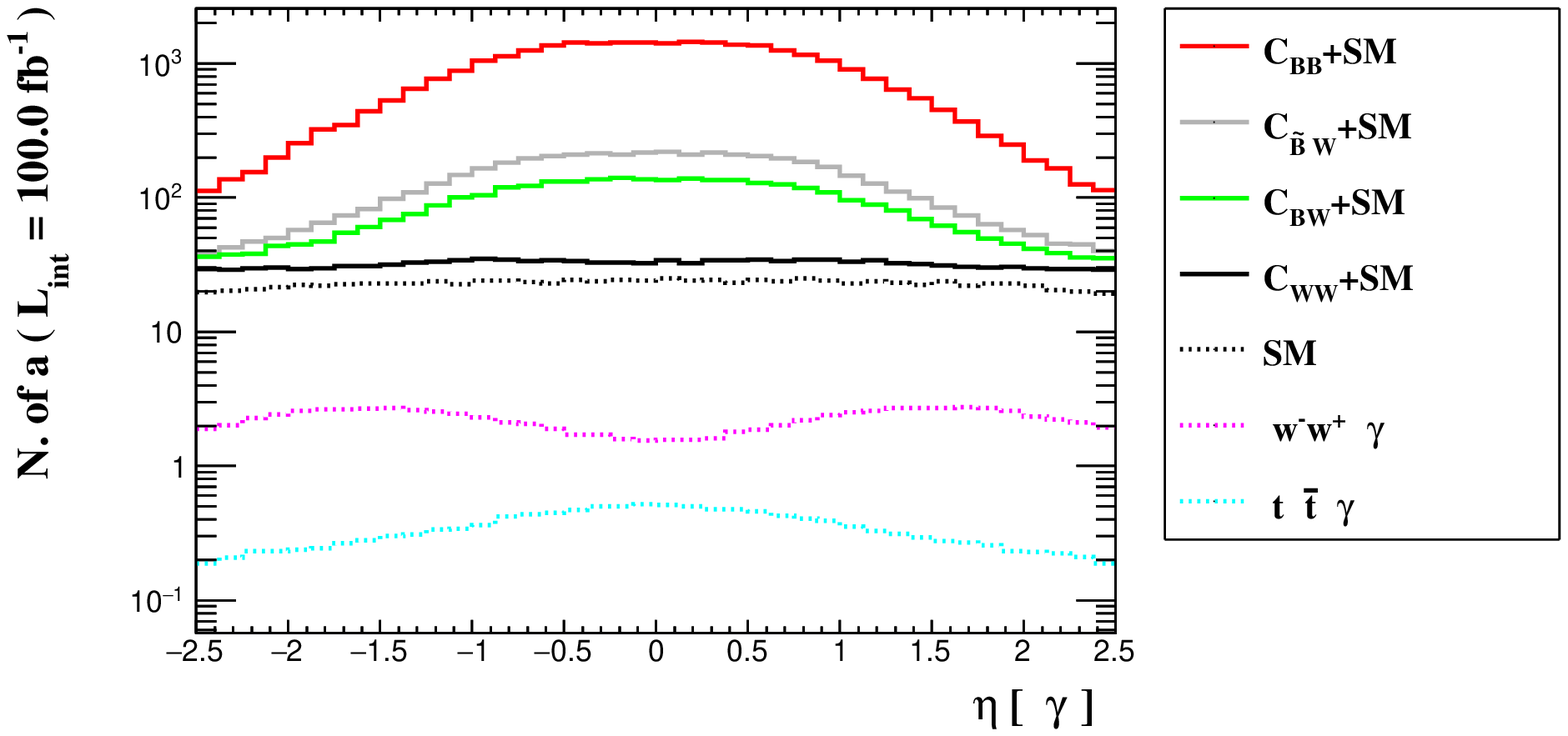}}}
\caption{The number of expected events as a function of $\eta_{\gamma}$ for the $e^+e^- \to Z\gamma$ signal and relevant backgrounds. In this figure, we have taken a value of  $1\hspace{1mm}{\rm TeV}^{-4}$ for each anomalous coupling.}
\label{Fig.4}
\end{figure}

All signal and background events analyses are simulated through UFO model file into {\sc MadGraph5}$\_$aMC@NLO \cite{Alwall:2014cvc} with approximately $1\times10^6$ events for each. It is necessary to apply some analysis cuts to distinguish the signal from the relevant backgrounds. For these cuts, we can use pseudo-rapidity $\eta^{\gamma}$, transverse momentum $p^\gamma_T$ and missing energy transverse $\slashed{E}_T$ for the photon in the final state of the process $e^-e^+\,\rightarrow\,\nu\bar{\nu}\gamma$. As seen in Fig.~\ref{Fig.2}, the transverse energy distribution of four different coupling signals and three different backgrounds is considered for the final state of the $e^-e^+\,\rightarrow\,\nu\bar{\nu}\gamma$ process. It can be seen that the signals differ significantly from the backgrounds at values of about 300 GeV. On the other hand, the transverse momentum distribution of the photon seen in Fig.~\ref{Fig.3} show significant deviations from the backgrounds for all couplings at values around 300 GeV. Analysis of the $\nu\bar{\nu}\gamma$ channel using the photon $p_T^{\gamma}$ distribution and the missing transverse energy distribution has the potential advantage since the branching ratio of the $Z$ boson decaying to the neutrino pair is greater than that of the lepton pair. There is no contribution from the final state bremsstrahlung and virtual photon to the $\nu\bar{\nu}\gamma$ channel \cite{Senol:2018gvg}. In addition, it is revealed in Fig.~\ref{Fig.4} that there is a differentiation between the signal for $C_{\widetilde{B}W}/{\Lambda^4}$, $C_{BB}/{\Lambda^4}$, $C_{BW}/{\Lambda^4}$ and $C_{WW}/{\Lambda^4}$ couplings and backgrounds in the distribution of the photon pseudo-rapidity in the range of $\pm$2.5. This signal results in Figs.~\ref{Fig.2}-\ref{Fig.4} are shown using 1 TeV$^{-4}$ for each anomalous coupling at the integrated luminosities of 100 fb$^{-1}$. Therefore, the cuts should be highlighted to adequately separate the signal from the SM, $W^-W^+\gamma$, and $t\bar{t}\gamma$ backgrounds in this study. The applied cut ﬂow in the analysis is summarized by the labels Cut-1, Cut-2 and Cut-3 in Table \ref{tab1}. After the cut flow, the number of events of the first two anomalous couplings giving the largest signal at $C_{BB}/{\Lambda^4}=C_{\widetilde{B}W}/{\Lambda^4}=1$ TeV$^{-4}$ and the relevant backgrounds are given in Table \ref{tab2} to examine the effects of applied cuts. It is seen that the cut flow suppresses the relevant backgrounds and the signals are very high relative to these backgrounds.

\begin{table}
\caption{Number of expected events for the process $e^+e^- \to Z\gamma$, the SM and relevant backgrounds processes after applied cuts given in Table I.}
\label{tab2}
\begin{tabular}{|c|c|c|c|c|c|c}
\hline
Kinematic cuts & $C_{BB}/\Lambda^{4}$ = 1 TeV$^{-4}$ & $C_{\widetilde{B}W}/\Lambda^{4}$=1 TeV$^{-4}$ &SM & $e^{-}e^{+}\to W^{-} W^{+}\gamma$ & $e^{-}e^{+}\to t \bar{t} \gamma $  \\
\hline
\hline
Cut-0 &30600 & 5093 & 915 & 89 & 14 \\
\hline
Cut-1  & 30600 & 5093 & 915 &  89 & 14  \\
\hline
Cut-2  & 30237 & 4957 & 823 &   31 & 5 \\
\hline
Cut-3  & 30237 & 4957 & 823 &  23 & 2 \\
\hline
\end{tabular}
\end{table}

It seems tempting that the collision of elementary particles such as electrons and positrons, accelerated at multi-TeV energies in the CLIC, directly tests new physics scenarios seeking answers to the open questions of the SM. The new physics research of the CLIC is built on two elements: Firstly, since leptons, which are elementary particles, are defined at a fundamental level, collisions in the lepton collider are clean without any hadronic activity and the measurements are very precise \cite{Gurkanli:2021ggq}. Second one, it is a discovery machine that can observe the new particle predicted in new physics models motivated by a large center-of-mass energy such as 3 TeV that the $e^-e^+$ collision can reach \cite{Franceschini:2020gfa}. For these reasons, the CLIC, which stands out with the discovery of new particles, stands out among the currently proposed $e^-e^+$ projects. The CLIC staging scenario assumes at the center-of-mass energy of 3 TeV and an integrated luminosity of 5 ab$^{-1}$. The CLIC experiment program specifies $\pm80\%$ longitudinal polarization for the electron beam and no polarized positron beam \cite{Roloff:2018tvb}. It is assumed that the weight of each event shares the luminosity in a ratio of 4:1 between the negative and positive polarization of the electron beam, with the result that the integrated luminosities are evaluated with ${\cal L}_{\text{int}}=4$ ab$^{-1}$ for $\mathcal{P}(e^-)=-80\%$ and ${\cal L}_{\text{int}}=1$ ab$^{-1}$ for $\mathcal{P}(e^-)=+80\%$ \cite{Weber:2020ymp}. Beam polarizations increase analysis capability and reduce systematic errors. It reveals new processes by enhancing the signal or suppressing the SM processes. The use of polarized electron beams helps in increasing signal rates and minimizing unwanted background processes \cite{Gurkanli:2021ggq,Spor:2021tgt}.

Using the polarized electron beam $\mathcal{P}(e^-)$ and the polarized positron beam $\mathcal{P}(e^+)$ in an $e^-e^+$ collider, the cross-section for a given process is obtained as follows, with respect to the four possible chiral cross-sections \cite{Fujii:2018ujn}:

\begin{eqnarray}
\label{eq.23} 
\sigma(\mathcal{P}_{e^-},\mathcal{P}_{e^+})=& \dfrac{1}{4} \Big\{(1+\mathcal{P}_{e^-})(1+\mathcal{P}_{e^+})\sigma_{RR}+(1-\mathcal{P}_{e^-})(1-\mathcal{P}_{e^+})\sigma_{LL} \\ \nonumber
&+(1+\mathcal{P}_{e^-})(1-\mathcal{P}_{e^+})\sigma_{RL}+(1-\mathcal{P}_{e^-})(1+\mathcal{P}_{e^+})\sigma_{LR}\Big\},
\end{eqnarray}

{\raggedright where $\sigma_{LR}$ presents the cross-section for left-handed polarized electron beam and right-handed polarized positron beam. Other cross-sections, $\sigma_{RL}$, $\sigma_{LL}$, and $\sigma_{RR}$, are defined in accordance with this definition. The unpolarized cross-section is written by}

\begin{eqnarray}
\label{eq.24} 
\sigma_0=\frac{1}{4} \{\sigma_{RR}+\sigma_{LL}+\sigma_{RL}+\sigma_{LR}\}.
\end{eqnarray}

\begin{figure}[t]
\centerline{\scalebox{1.3}{\includegraphics{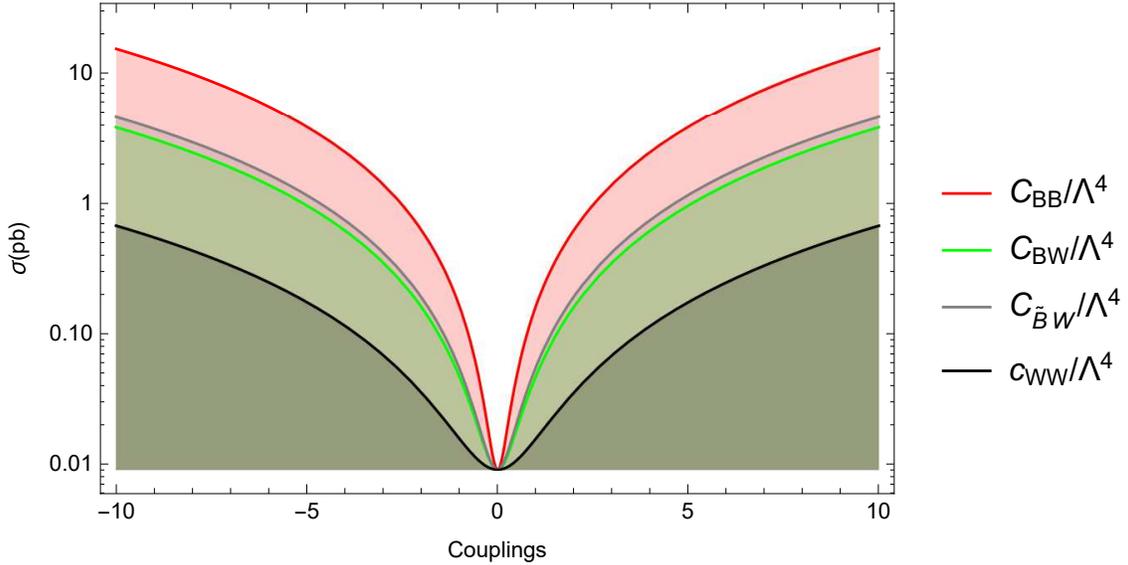}}}
\caption{Production cross-section for the process $e^+e^-\to Z\gamma$ in terms of the anomalous $C_{BB}/\Lambda^{4}$, $C_{BW}/\Lambda^{4}$, $C_{\widetilde{B}W}/\Lambda^{4}$, $C_{WW}/\Lambda^{4}$ couplings for the CLIC with $\sqrt{s}=3$ TeV and $P_{e^-}=-80\%$.}
\label{Fig.5}
\end{figure}

\begin{figure}[t]
\centerline{\scalebox{1.3}{\includegraphics{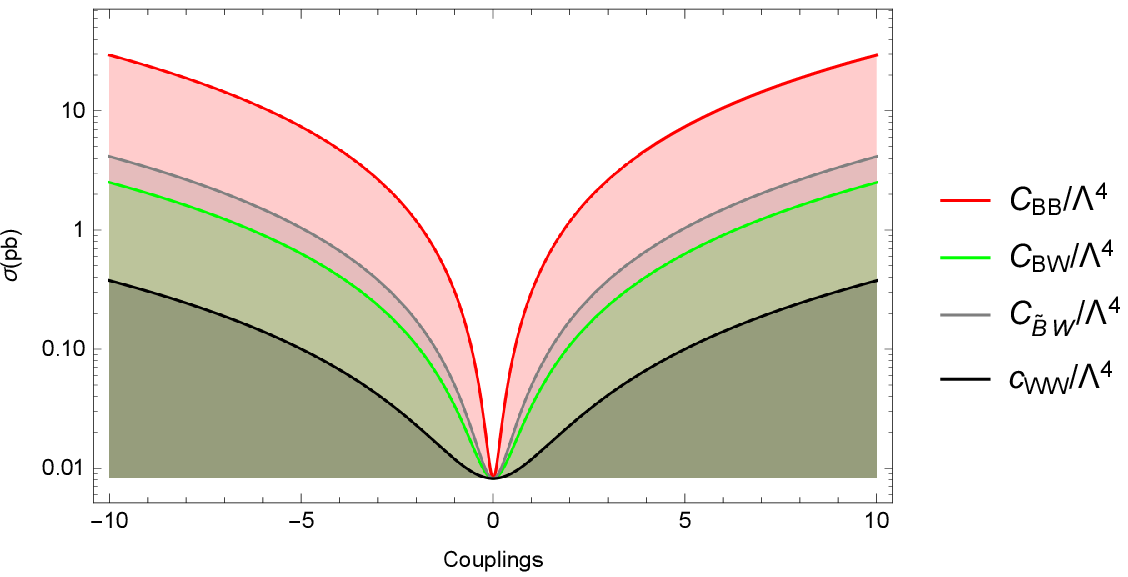}}}
\caption{Same as Fig.5 but for unpolarized beams $P_{e^-}=0\%$.}
\label{Fig.6}
\end{figure}

\begin{figure}[t]
\centerline{\scalebox{1.3}{\includegraphics{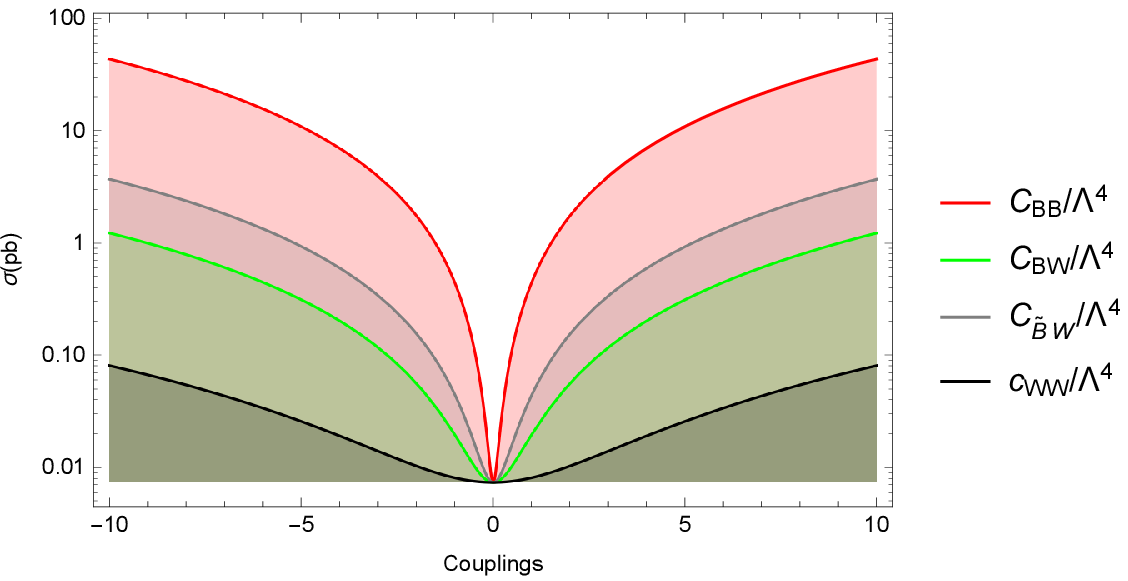}}}
\caption{Same as Fig.5 but for polarized beams $P_{e^-}=80\%$.}
\label{Fig.7}
\end{figure}

Total cross-sections of the process $e^-e^+\,\rightarrow\,Z\gamma$ as a function of anomalous $C_{\widetilde{B}W}/{\Lambda^4}$, $C_{BB}/{\Lambda^4}$, $C_{BW}/{\Lambda^4}$ and $C_{WW}/{\Lambda^4}$ couplings, in the CLIC with $\sqrt{s}=3$ TeV are given in Fig.~\ref{Fig.5} for an electron beam polarization of -80$\%$, in Fig.~\ref{Fig.6} for unpolarized beams, and in Fig.~\ref{Fig.7} for an electron beam polarization of +80$\%$. The other three couplings are fixed to zero to find the variation of the total cross-section with respect to the function of each anomalous coupling. In these total cross-section calculations, the cuts in Table \ref{tab1} are used to suppress the backgrounds. If the total cross-sections according to the coupling values of each anomalous coupling are ordered from the highest to the lowest, it can be seen from Figs.~\ref{Fig.5}-\ref{Fig.7} that they are $C_{BB}/{\Lambda^4}$, $C_{\widetilde{B}W}/{\Lambda^4}$, $C_{BW}/{\Lambda^4}$ and $C_{WW}/{\Lambda^4}$.

\section{SENSITIVITIES ON THE ANOMALOUS COUPLINGS}

The sensitivities of the anomalous $C_{\widetilde{B}W}/{\Lambda^4}$, $C_{BB}/{\Lambda^4}$, $C_{BW}/{\Lambda^4}$ and $C_{WW}/{\Lambda^4}$ couplings in process $e^-e^+\,\rightarrow\,Z\gamma$ are obtained by using a $\chi^2$ method with systematic errors at 95$\%$ C.L., defined by

\begin{eqnarray}
\label{eq.25} 
\chi^2=\left(\frac{\sigma_{SM}-\sigma_{NP}}{\sigma_{SM}\sqrt{\left(\delta_{st}\right)^2+\left(\delta_{sys}\right)^2}}\right)^2
\end{eqnarray}

{\raggedright where $\sigma_{SM}$ is the cross-section of relevant SM backgrounds and $\sigma_{NP}$ is the total cross-section containing contributions from the presence of both new physics beyond the SM and relevant SM backgrounds. $\delta_{st}=\frac{1}{\sqrt {N_{SM}}}$ and $\delta_{sys}$ are the statistical error and the systematic error, respectively. The number of events of relevant SM backgrounds is given with $N_{SM}={\cal L}_{\text{int}} \times \sigma_{SM}$, where ${\cal L}_{\text{int}}$ is the integrated luminosity.}

In the analyses of the studies, there are systematic uncertainties related to the measurement of the cross-sections containing neutrino production. In general, systematic uncertainties arising from many reasons such as detector luminosity, trigger efficiencies, jet energy calibration, bjet tagging efficiencies, lepton identification, backgrounds, initial and final state radiation, parton distribution functions should be included in statistical methods \cite{Khoriauli:2008bnm}. The process $pp\,\rightarrow\,\nu \bar{\nu} \gamma$ including the anomalous $ZZ\gamma$ and $Z\gamma\gamma$ couplings is also examined for the FCC-hh with same systematic errors \cite{Senol:2018gvg}. In the study on the anomalous magnetic and electric dipole moment of the neutrino, the systematic uncertainty for the process $pp\,\rightarrow\,\nu\bar{\nu}\gamma$ at the LHC is considered to be 0$\%$, 5$\%$ and 10$\%$ \cite{Rodriguez:2019ygz}. In a similar study performed on the CLIC for the $e^-e^+\,\rightarrow\,\nu\bar{\nu}\gamma$ process, the systematic uncertainty is again considered as 0$\%$, 5$\%$ and 10$\%$ \cite{Rodriguez:2018okh}. For this reason, taking into account the previous study, we choose the systematic uncertainties $\delta_{sys}=0\%$, 5$\%$ and 10$\%$ for the CLIC.

\begin{table}
\caption{Constraints on aNTGCs $C_{BB}/\Lambda^{4}$, $C_{BW}/\Lambda^{4}$, $C_{\widetilde{B}W}/\Lambda^{4}$ and $C_{WW}/\Lambda^{4}$ via $e^+e^- \to Z\gamma$ process at the CLIC. }
\label{tab3}
\begin{tabular}{|c|c|c|c|c|}
\hline
$P_{e^-}$              &      & $0\%$       & $-80\%$    & $80\%$ \\
\hline
Couplings (TeV$^{-4}$) & & ${\cal L}=5$ ab$^{-1}$ & ${\cal L}=4$ ab$^{-1}$ & ${\cal L}=1$ ab$^{-1}$ \\
\hline
                      &$\delta=0\%$       &$[-1.35;2.00]\times10^{-2}$  &$[-2.19;2.81]\times10^{-2}$ &$[-1.66;2.34]\times10^{-2}$ \\
$C_{BB}/\Lambda^{4}$  &$\delta=5\%$       &$[-4.95;5.61]\times10^{-2}$  &$[-7.40;8.00]\times10^{-2}$ &$[-3.81;4.49]\times10^{-2}$ \\
                      &$\ \, \delta=10\%$ &$[-7.11;7.77]\times10^{-2}$  &$[-1.06;1.11]\times10^{-1}$ &$[-5.47;6.15]\times10^{-2}$ \\
\hline
                      &$\delta=0\%$       &$[-5.38;5.94]\times10^{-2}$  &$[-4.75;5.24]\times10^{-2}$ &$[-1.15;1.21]\times10^{-1}$ \\
$C_{BW}/\Lambda^{4}$  &$\delta=5\%$       &$[-1.78;1.83]\times10^{-1}$  &$[-1.52;1.57]\times10^{-1}$ &$[-2.44;2.50]\times10^{-1}$ \\
                      &$\ \, \delta=10\%$ &$[-2.52; 2.58]\times10^{-1}$ &$[-2.16;2.21]\times10^{-1}$ &$[-3.43;3.49]\times10^{-1}$ \\
\hline
                      &$\delta=0\%$       &$[-4.28;4.52]\times10^{-2}$  &$[-4.97;4.16]\times10^{-2}$ &$[-6.02;7.62] \times10^{-2} $ \\
$C_{\widetilde{B}W}/\Lambda^{4}$  &$\delta=5\%$       &$[-1.39;1.42]\times10^{-1}$  &$[-1.45;1.37]\times10^{-1}$ &$[-1.34;1.50] \times10^{-1}$ \\
                      &$\ \, \delta=10\%$ &$[-1.97; 1.99]\times10^{-1}$ &$[-2.03;1.95]\times10^{-1}$ &$[-1.91;2.07] \times10^{-1}$ \\
\hline
                      &$\delta=0\%$       &$[-1.45;1.50]\times10^{-1}$  &$[-1.17;1.12]\times10^{-1}$ &$[-4.79;4.80]\times10^{-1}$ \\
$C_{WW}/\Lambda^{4}$  &$\delta=5\%$       &$[-4.68;4.73]\times10^{-1}$  &$[-3.67;3.73]\times10^{-1}$ &$[-1.01;1.01]$ \\
                      &$\ \, \delta=10\%$ &$[-6.61; 6.67]\times10^{-1}$ &$[-5.20;5.25]\times10^{-1}$ &$[-1.41;1.41]$ \\
\hline
\end{tabular}
\end{table}

The 95$\%$ C.L. limits of anomalous $C_{BB}/{\Lambda^4}$, $C_{\widetilde{B}W}/{\Lambda^4}$, $C_{BW}/{\Lambda^4}$ and $C_{WW}/{\Lambda^4}$ couplings through process $e^-e^+\,\rightarrow\,Z\gamma$ at the CLIC are investigated in Table \ref{tab3} with and without systematic error according to configurations of beam polarization and integrated luminosity. If we first compare the limits between four anomalous couplings, the highest sensitivity for all three configurations belongs to the anomalous $C_{BB}/{\Lambda^4}$ coupling. Second, if configurations of beam polarization and integrated luminosity are evaluated, anomalous $C_{BB}/{\Lambda^4}$ and $C_{\widetilde{B}W}/{\Lambda^4}$ couplings are most sensitive in $\mathcal{P}(e^-)=0\%$ and ${\cal L}_{\text{int}}=5$ ab$^{-1}$ configuration, while anomalous $C_{BW}/{\Lambda^4}$ and $C_{WW}/{\Lambda^4}$ couplings are most sensitive in $\mathcal{P}(e^-)=-80\%$ and ${\cal L}_{\text{int}}=4$ ab$^{-1}$ configuration. This shows that at the limits of anomalous $C_{BW}/{\Lambda^4}$ and $C_{WW}/{\Lambda^4}$ couplings, beam polarization has a sensitivity-enhancing effect. The most sensitive limits of the aNTGCs without systematic error are as follows, respectively:

\begin{eqnarray}
\label{eq.26} 
C_{BB}/{\Lambda^4}=[-1.35; 2.00]\times10^{-2}\,\text{TeV}^{-4}\,,
\end{eqnarray}
\begin{eqnarray}
\label{eq.27} 
C_{\widetilde{B}W}/{\Lambda^4}=[-4.28; 4.52]\times10^{-2}\,\text{TeV}^{-4}\,,
\end{eqnarray}
\begin{eqnarray}
\label{eq.28} 
C_{BW}/{\Lambda^4}=[-4.75; 5.24]\times10^{-2}\,\text{TeV}^{-4}\,,
\end{eqnarray}
\begin{eqnarray}
\label{eq.29} 
C_{WW}/{\Lambda^4}=[-1.17; 1.12]\times10^{-1}\,\text{TeV}^{-4}\,.
\end{eqnarray}

\begin{figure}[t]
\centerline{\scalebox{0.75}{\includegraphics{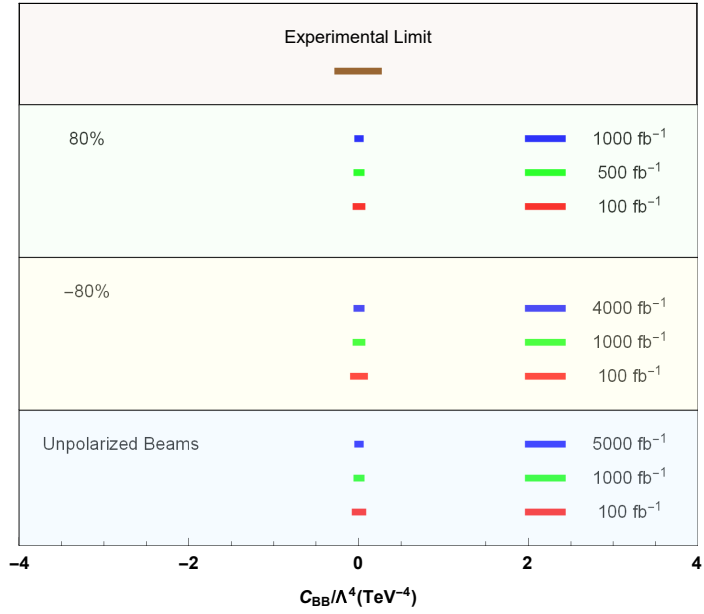}}}
\caption{Comparison of the current experimental limits and projected sensitivity on the anomalous $C_{BB}/\Lambda^4$ coupling for expected luminosities of ${\cal L}_{\text{int}}=100, 500, 1000, 4000, 5000\hspace{0.8mm}$ fb$^{-1}$ and $\sqrt{s}=3\hspace{0.8mm}{\rm TeV}$ at the CLIC. We consider $P_{e^-}=-80\%, 0\%, 80\%$.}
\label{Fig.8}
\end{figure}

\begin{figure}[t]
\centerline{\scalebox{0.75}{\includegraphics{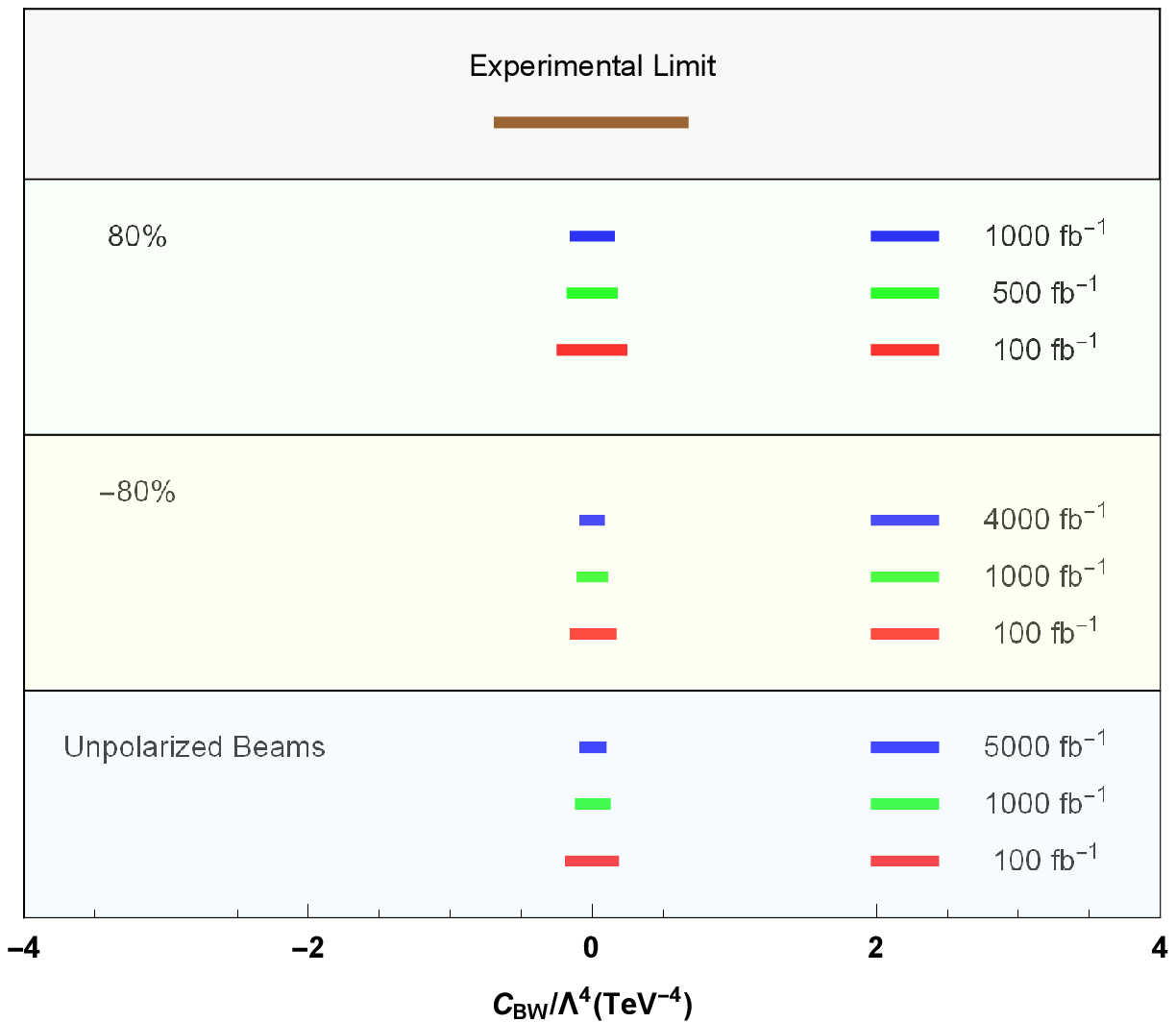}}}
\caption{Same as Fig.~\ref{Fig.8} but for the anomalous $C_{BW}/\Lambda^4$} coupling.
\label{Fig.9}
\end{figure}

\begin{figure}[t]
\centerline{\scalebox{0.75}{\includegraphics{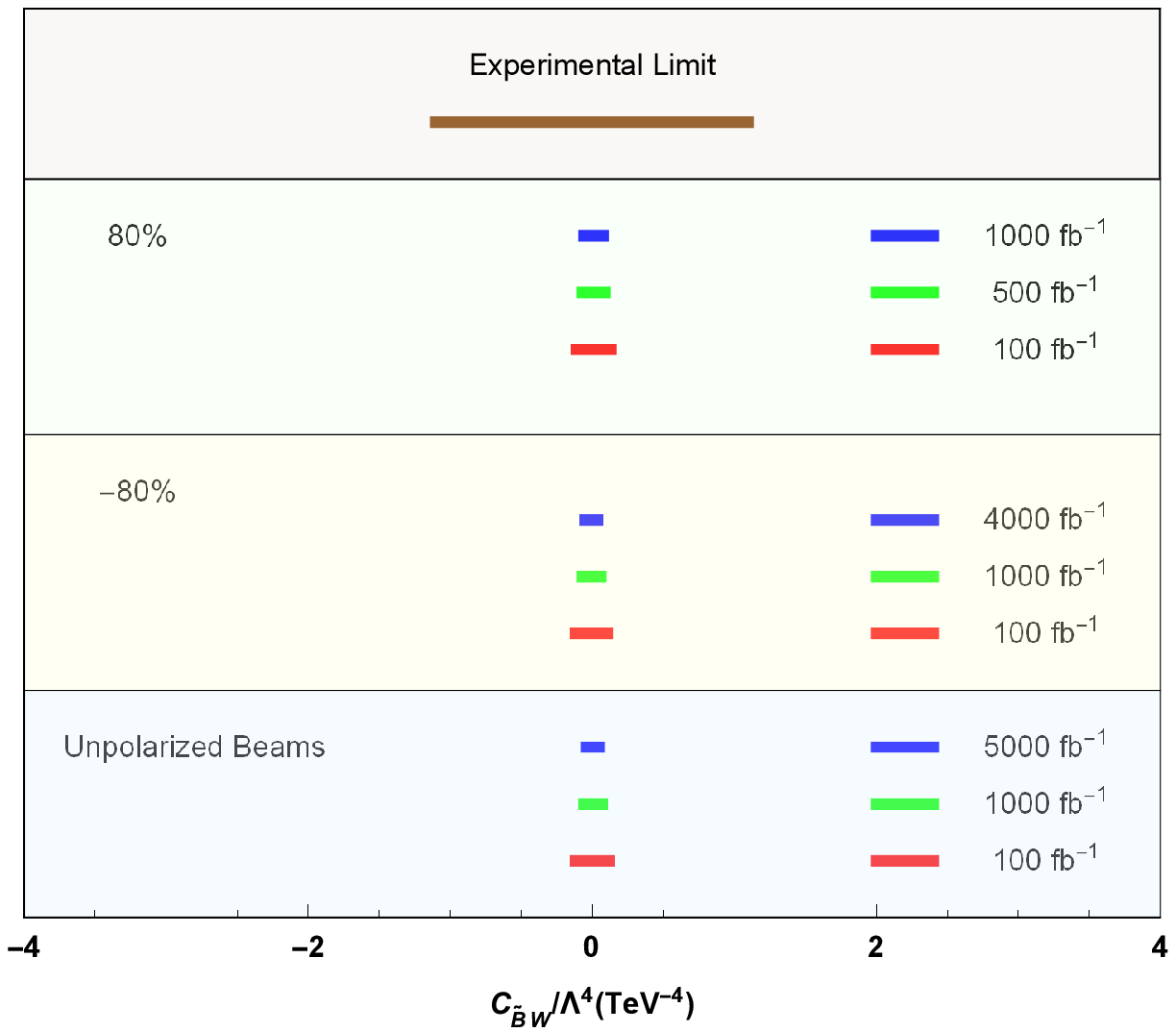}}}
\caption{Same as Fig.~\ref{Fig.8} but for the anomalous $C_{\widetilde{B}W}/\Lambda^4$} coupling.
\label{Fig.10}
\end{figure}

\begin{figure}[t]
\centerline{\scalebox{0.75}{\includegraphics{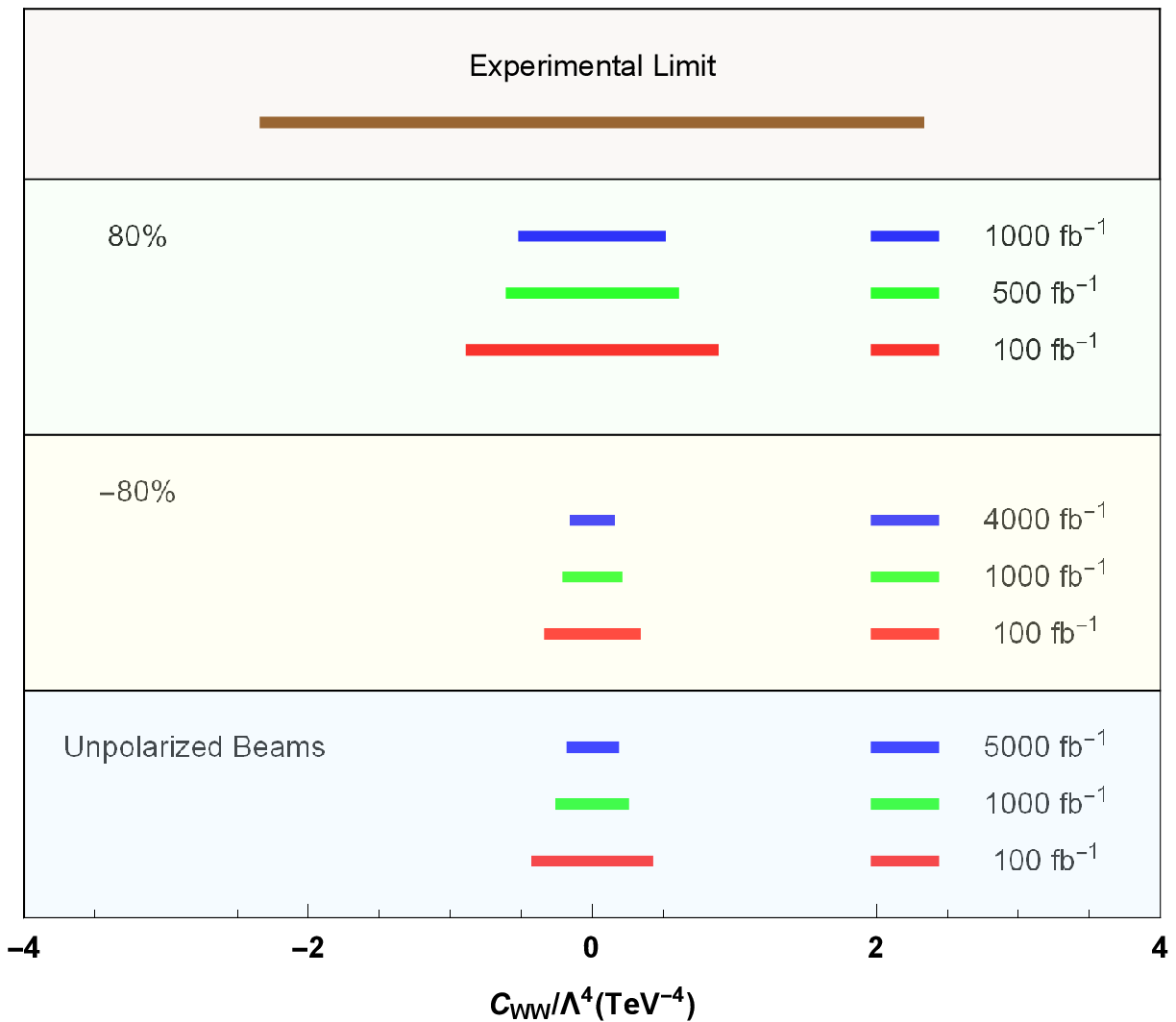}}}
\caption{Same as Fig.~\ref{Fig.8} but for the anomalous $C_{WW}/\Lambda^4$} coupling.
\label{Fig.11}
\end{figure}

Figs.~\ref{Fig.8}-\ref{Fig.11} summarize the sensitivities of anomalous $C_{BB}/{\Lambda^4}$, $C_{BW}/{\Lambda^4}$, $C_{\widetilde{B}W}/{\Lambda^4}$ and $C_{WW}/{\Lambda^4}$ couplings according to beam polarization and integrated luminosities, respectively, at the CLIC with $\sqrt{s}=3$ TeV, by comparison with the current experimental limits of the ATLAS Collaboration \cite{Aaboud:2018ybz}. The integrated luminosity values according to the beam polarization type, which are mentioned in Section~\ref{Sec2}, are reached in three steps starting from 100 fb$^{-1}$ and the change of anomalous couplings with this gradual luminosity increase in each polarization type are investigated. Most importantly, we have obtained limits with better sensitivity than current experimental limits for all anomalous couplings.

According to the EFT approach, the minimum coupling value of the coefficients is required to put the operator scale $\Lambda$ beyond the reach of the kinematic range of the distributions. The new physics characteristic scale $\Lambda$ can be related to the coefficients of dimension-eight operators and an upper bound can be placed on this scale \cite{Senol:2020hbh}. For $C=O(1)$ couplings, we find $\Lambda<\sqrt{2\pi \upsilon \sqrt{s}}\sim3.04$ TeV.

\section{Conclusions}

In this study, we carried out the process $e^{-} e^{+}\to Z\gamma \to (\nu \nu) \gamma $ to probe the $Z\gamma\gamma$ and $ZZ\gamma$ aNTGC at CLIC. In the analysis, cut based technique is applied to separate the signal and the relevant SM background. Besides, transverse momentum and the pseudo-rapidity of the final state photon, missing energy transverse are selected for the kinematic cuts. The effects of the selected cuts on the number of events for both signal and SM background have been given with a cut-flow chart. To finalize the study, we obtained the sensitivity of dimension-eight $CP$-conserving $C_{\widetilde{B}W}/{\Lambda^4}$ coupling and $CP$-violating $C_{BB}/{\Lambda^4}$, $C_{BW}/{\Lambda^4}$, $C_{WW}/{\Lambda^4}$ couplings with $95\%$ C.L. for unpolarized and $\mp{80\%}$ polarized electron beam that have recomposed for systematic uncertainties of $0\%$, $5\%$ and $10\%$ at $\sqrt{s}=3$ TeV. We used the integrated luminosities of ${\cal L}_{\text{int}}=5$ $\rm ab^{-1}$, ${\cal L}_{\text{int}}=4$ $\rm ab^{-1}$ and ${\cal L}_{\text{int}}=1$ $\rm ab^{-1}$ to calculate sensitivities for unpolarized and $\mp{80\%}$ polarized electron beams, respectively. Our results improve the sensitivities of the aNTGC couplings with respect to the latest experimental results and phenomenological studies by a factor of $10$ to $30$ times.

\end{document}